\documentclass [preprint,showkeys,author-year] {jasatex}

\usepackage{amsmath}
\usepackage{graphics}
\usepackage{inputenc}
\newcommand{\p}{\partial}
\newcommand{\be}{\begin{equation}}
\newcommand{\ee}{\end{equation}}
\newcommand{\rhot}{\tilde{\rho}}
\newcommand{\ba}{\boldsymbol}
\newcommand{\di}{\displaystyle}
\DeclareTextSymbol{\degre}{OT1}{23}
\begin{document}
\preprint{AIP/123-QED}
%--- Pour le titre ---
\title[Reflectivity perturbations in porous media]{Perturbations of the seismic reflectivity of a fluid-saturated depth-dependent poro-elastic medium}
\author{Louis De Barros}
\email{louis.debarros@ujf-grenoble.fr}
\author{Michel Dietrich}
\altaffiliation[Now at ]{Institut Fran\c cais du P\'etrole, 92852 Rueil-Malmaison, France }
\affiliation{Laboratoire de G\'eophysique Interne et Tectonophysique (LGIT), CNRS, Universit\'e Joseph Fourier, BP 53, 38041 Grenoble Cedex 9, France}
\date{\today}
\begin{abstract}
Analytical formulas are derived to compute the first-order effects produced by plane inhomogeneities on the point source seismic response of a fluid-filled stratified porous medium. 
The derivation is achieved by a perturbation analysis of the poro-elastic wave equations in the plane-wave domain using the Born approximation. 
This approach yields the Fr\'echet derivatives of the $P-SV$- and $SH$-wave responses in terms of the Green's functions of the unperturbed medium.
The accuracy and stability of the derived operators are checked by comparing, in the time-distance domain, differential seismograms computed from these analytical expressions with complete solutions obtained by introducing discrete perturbations into the model properties. 
For vertical and horizontal point forces, it is found that the Fr\'echet derivative approach is remarkably accurate for small and localized  perturbations of the medium properties which are consistent with the Born approximation requirements. 
Furthermore, the first-order formulation appears to be stable at all source-receiver offsets.
The porosity, consolidation parameter, solid density and mineral shear modulus emerge as the most sensitive parameters in forward and inverse modeling problems. 
Finally, the Amplitude-Versus-Angle response of a thin layer shows strong coupling effects between several model parameters. 
\end{abstract}


%========================================= Corps ==============================================

\maketitle              % écrit le titre

%-------------------------------------- Introduction ------------------------------------------

\section{Introduction}

The evaluation of the sensitivity of a seismic wave field to small perturbations of the material properties is a classical issue of seismology which arises in the solution of forward and inverse scattering problems \citep{aki80, tarantola84a}. 
Sensitivity operators are mainly useful for the optimal design of field or laboratory experiments, for the interpretation of time-lapse monitoring surveys, and for the development of imaging and linearized inversion techniques.
In particular, sensitivity operators play a central role in least-square inversion schemes using gradient techniques.
All these applications, especially the latter, call for fast and effective numerical computation methods of the sensitivity operators.
Indeed, the most intuitive approach to compute the perturbational or differential seismograms in a structure described by $N$ parameters is to use a finite perturbation scheme that demands $N+1$ forward modeling computations.
This approach rapidly becomes prohibitive as the number of parameters increases.
The problem can be solved more elegantly by first deriving the so-called Fr\'echet derivatives of the seismic wave fields with respect to material properties.
Although restricted to first-order effects only, this procedure makes it possible to efficiently predict small changes of the seismic response resulting from slight modifications of the material properties.
Furthermore, this solution can be implemented by solving only one large forward problem.

One of the most significant contributions in this area is the work of %Tarantola 
\citet{tarantola84a} who applied a first-order perturbation analysis to the elastodynamic wave equation to derive a series of general formulas relating the scattered wave field to heterogeneities in an arbitrarily complex elastic medium. 
The same approach was subsequently used in a more restrictive sense for layered media in the plane-wave domain by \citet{Pan88}  in the acoustic case and by %Dietrich and Kormendi 
\citet{dietrich90} in the $P-SV$ case. 
The Fr\'echet derivatives obtained in these different cases are all expressed as combinations of the incident wave field generated by the seismic source at the scatterer location and the Green's functions between the scatterer and receiver locations.
The structure of these expressions underlines the fact that the scattered waves due to perturbations in the material properties (usually density and acoustic or elastic parameters) can be interpreted as a wave field generated by a set of secondary body forces coincident with the heterogeneities, and determined by complex interactions between the incident waves and the medium perturbations.
This structure is met with all problems of acoustic, seismic or electromagnetic wave propagation in weakly inhomogeneous media, and will also be found in more complex situations such as anisotropic or poro-elastic media.

The poro-elastic model, which is the subject of this paper, involves more parameters than the visco-elastic case, but on the other hand, the wave velocities, attenuation and dispersion characteristics are computed from the medium's intrinsic properties without having to resort to empirical relationships. 
Since the pioneering work of %Biot
\citet{biot56}, many authors \citep[for example]{dutta79a, auriault85, johnson94} have introduced improvements of the poro-elastodynamic equations, either by averaging or by integrating techniques. The \citet{biot56} theory and its applications is still a field of active research, as demonstrated by the large number of current publications devoted to the subject (see, e.g., Trifunac, 2006\nocite{trifunac06}).
The forward problem, i.e., the computation of synthetic seismograms in poro-elastic media has been solved in different configurations and with several techniques \citep{dai95, carcione96, haartsen97, garambois02}. 
However, the inverse problem has only been rarely addressed, and to our knowledge, it has never been without first estimating the wave velocities \citep{chotiros02, berryman02, spikes06}.
Yet, inversion algorithms can provide useful information on the material properties, notably
permeability and porosity which are the most important parameters to characterize porous media.

The main objective of this work is to extend the methodology used in the elastic case \citep{dietrich90} to obtain the Fr\'echet derivatives for stratified poro-elastic media. 
We consider here a depth-dependent, fluid-saturated porous medium representing reservoir rocks or sedimentary layers. 
The computation of the point source seismic response of the layered structure is carried out by combining the Generalized Reflection and Transmission Matrix Method \citep{kennett83} with the discrete wavenumber technique \citep{bouchon81}.
This combination was already implemented by \citet{garambois02} for the numerical simulation of the coupled seismic and electromagnetic wave propagation in porous media by using the work of \citet{pride94}. 
In the following sections, we first present the governing equations and constitutive parameters for porous materials before expressing the wave propagation equations for depth-dependent media. 
Next, we develop the analytical computation of the Fr\'echet derivatives in the fre\-quency--ray parameter domain for the $P-SV$- and $SH$-wave cases. 
Finally, we check the accuracy of the sensitivity operators obtained in the time-distance domain, both in an infinite medium and in a complex seismic model. 
We conclude with the sensitivity of the seismic waveforms with respect to the different model parameters.

\section{Wave propagation in stratified porous media}

\subsection{Governing equations}

Assuming a $e^{-i \omega t}$ dependence, where $t$ is time and $\omega$ the angular frequency, \citet{pride94, pride03} rewrote Biot's (\citeyear{biot56}) equations of poroelasticity in the form
\be \label{eq201}
\left\{ \begin{array}{rcl} 
\vspace{3pt}
\nabla\cdot{ \ba{\tau}}& = & -\omega^2(\rho \,{\bf u}+\rho_{f} \,{\bf w}) \\
\vspace{3pt}
{\ba \tau}& = & [~K_U\,\nabla\cdot{\bf u} +C\,\nabla\cdot{\bf w}~]\,{\bf I} + G\,[~\nabla {\bf u}+(\nabla {\bf u})^T - 2/3\,(\nabla\cdot({\bf u} \, {\bf I}) )~] \\
\vspace{3pt}
-P & = & C\,\nabla\cdot{\bf u} + M\,\nabla\cdot{\bf w} \\
-\nabla P & = & -\omega^2 \, \rho_{f} \, {\bf u} -\omega^2 \rhot \, {\bf w} ~, 
\end{array} \right.
\ee
where ${\bf u}$ and ${\bf w}$ are the solid average displacement and the relative fluid-to-solid displacement, respectively.
More precisely, defining $\bf u_s$ and $\bf u_f$ as the displacements of the solid and fluid phases of a porous continuum, we can write $\bf u \simeq \bf u_s$ and $\bf w = \phi (\bf u_s-\bf u_f)$, where $\phi$ is the porosity.
$P$ represents the interstitial pressure and $\ba{\tau}$ is the 3$\times$3 stress tensor.
$\rho$ denotes the density of the porous medium which is related to the fluid density $\rho_f$, solid density $\rho_s$ and porosity $\phi$ via the relationship
\be \label{eq202}
\rho=(1-\phi)\, \rho_s + \phi\, \rho_f ~.
\ee
The undrained bulk modulus $K_U$ is defined under the condition $\bf w=0$. 
$G$ is the shear modulus of a drained or an undrained medium as it is independent of the fluid properties \citep{gassmann51}. 
The fluid storage coefficient $M$ represents the amount of fluid a sample can accumulate at constant sample volume. 
Biot's $C$ modulus is a mechanical parameter describing the variation of the fluid pressure due to a change of the sample volume in an undrained medium.
At low frequencies, these parameters as well as the Lam\'e parameter $\lambda_u$ defined below are real, frequency-independent and can be expressed in terms of the drained modulus $K_D$, porosity $\phi$, mineral modulus of the grains $K_s$ and fluid modulus $K_f$ \citep{gassmann51}:
\begin{eqnarray} \label{eq203}
K_U & = &  \frac{\phi\, K_D+\left[~1-(1+\phi)\,\di{\frac{K_D}{K_s}}~\right] K_f}{\phi \,(1+\Delta)}~, \hspace{10pt} \lambda_U=K_U -\frac{2}{3}\,G \nonumber\\
C & = & \frac{\left[~1-\di{\frac{K_D}{K_s}}~\right]\,K_f}{\phi \,(1+\Delta)}~, \hspace{10pt} 
M =\frac{ K_f}{\phi \,(1+\Delta)} \\
\hspace{-10pt}\textrm{with}\hspace{10pt}    
\Delta & = & \frac{1-\phi}{\phi} \,\frac{K_f}{K_s} \left[~1- \frac{K_D}{(1-\phi) \, K_s}~\right] ~. \nonumber
\end{eqnarray}

It is also possible to link the bulk properties $K_D$ and $G$ to the porosity and constitutive mineral properties via empirical relationships derived from experimental results \citep{pride03,bemer04}:
\be \label{eq205}
K_D = K_s \,\frac{1-\phi}{1+{c_s}\,\phi} \qquad \textrm{and} \qquad  G = G_s \,\frac{1-\phi}{1+3 \,{c_s} \,\phi /2} ~.
\ee
Equations (\ref{eq205}) have the merit of being very simple and introduce only two additional parameters, namely, the shear modulus of the grains $G_s$ and the consolidation parameter $c_s$.
The latter mainly depends on the cementing properties of the grains, but also on the pore shape. 
The value of the consolidation parameter $c_s$ is typically between 2 to 20 in a consolidated medium, and can be very much greater than 20 in an unconsolidated soil. 

Finally, the wave attenuation is explained by Darcy's law which uses a complex, frequency-dependent dynamic permeability \citep{johnson94}: 
\be \label{eq206}
\rhot= i \frac{\eta}{\omega \, k(\omega)} \hspace{10pt}\textrm{with}\hspace{10pt}
k(\omega)= k_0 / \left[~\sqrt{1-i\,\frac{4}{n_J} \frac{\omega}{\omega_c}}-i\,\frac{\omega}{\omega_c}~\right] ~.
\ee
The dynamic permeability $k(\omega)$ tends toward the hydrogeological (dc) permeability $k_0$ at low frequencies where viscous losses are dominant.
It includes a correction accounting for the inertial effects at higher frequencies. 
These two domains are separated by the relaxation frequency
\be \label{eq207}
\omega_c= \frac{\eta}{{\rho_f} F {k_0}}
\ee
where $\eta$ is the viscosity of the fluid.
Archie's law ( $F=\phi^{-m}$ ) expresses the electrical formation factor $F$ in terms of the porosity $\phi$ and cementing exponent $m$ whose value is between 1 to 2 depending on the pore topology. 
Parameter $n_J$ is considered constant and equal to 8 to simplify the equations.
We refer the reader to the work of \citet{pride03} for more information on the parameters used in this study.

\subsection{Coupled second-order equations for plane waves}

The horizontally layered model lends itself to a number of analytical developments if one performs a plane wave decomposition of the wave fields represented by equations (\ref{eq201}).
This procedure involves a series of changes of variables and integral transforms which are described in detail by \citet{kennett83} in the elastic case. 
As in the elastic case, the introduction of new variables \citep{hudson69} leads to a useful decomposition into $P-SV$- and $SH$-wave systems in cylindrical coordinates. 
By applying the whole sequence of transformations and arranging the terms, we find that the governing equations of the $P-SV$-wave system in depth-dependent poro-elastic media reduce to the following system of second-order equations in the angular frequency $\omega$ and ray parameter $p$ domain:
\be \label{eq210}
\left\{ \begin{array}{rcl} 
F_{1z} & = & \di{\frac {\p }{\p z}}\,\left[~(
   \lambda_U  + 2\,G )\,\di{\frac {\p U}{\p z}} + C\,\di{\frac {\p W}{\p z}} - 
   \omega \,p\,(\lambda_U \,V + C\,X)~\right] - \omega \,p~G \, \di{\frac {\p V}{\p z}} \\
\vspace{6 pt}
& & + \,\omega ^{2}\,\left[~\rho\,U - p^{2}\,G \,U + {\rho_f}\,W~\right]\\
F_{1r} & = & \di{\frac {\p }{\p z}}\,\left[~G \,\di{\frac {\p V}{\p z}} + 
   \omega \,p~G \,U~\right] + \omega \,p\,\left[~\lambda_U\,\di{\frac {\p U}{\p z}} + 
   C\,\di{\frac {\p W}{\p z}}~\right]\\
\vspace{6pt}
 & & + \,\omega ^{2}\,\left[~{\rho}\,V - p^{2}\,(\lambda_U  + 
   2\,G)\,V - p^{2}\,C\,X~\right]\\
F_{2z} & = & \di{\frac {\p }{\p z}}\,\left[~C\,\di{\frac {\p U}{\p z}} - 
   \omega \,p~C\,V + M\,\di{\frac {\p W}{\p z}} - \omega \,p\,M\,X~\right]
\vspace{6pt}
+ \omega ^{2}\,\left[~\rho_f\,U + \tilde{\rho}\,W~\right]\\
F_{2r} & = & \omega \,p\,\left[~C\,\di{\frac {\p U}{\p z}} + 
   M\,\di{\frac {\p W}{\p z}}~\right] + \omega^{2}\,\left[~\rho_f\,V + 
   \tilde{\rho}\,X - p^{2}\,(C\,V + M\,X)~\right]~.\\
\end{array} \right. 
\ee
In these equations, $U=U(z_R,\omega;z_S)$ and $V=V(z_R,\omega;z_S)$ respectively denote the vertical and radial components of the solid displacements.
Similarly, $W=W(z_R,\omega;z_S)$ and $X=X(z_R,\omega;z_S)$ respectively denote the vertical and radial components of the relative fluid-to-solid displacements.
Variables $z_R$ and  $z_S$ stand for the receiver and seismic source depths. 
Equations (\ref{eq210}) are valid in the presence of body forces ${\bf{F_1}}=[F_{1z}(z_S,\omega),F_{1r}(z_S,\omega)]^T$ and ${\bf{F_2}}=[F_{2z}(z_S,\omega),F_{2r}(z_S,\omega)]^T$ defined by their vertical ($z$ index) and radial ($r$ index) components: force $\bf{F_1}$ is applied on an average volume of porous medium and represents a stress discontinuity, while force $\bf{F_2}$ is derived from the pressure gradient in the fluid.

We can then cast equations (\ref{eq210}) in the form of a matrix differential equation as
\be \label{eq211}
{\bf L}^{^{\textrm{PSV}}} {\bf Q}^{^{\textrm{PSV}}} = {\bf F}^{^{\textrm{PSV}}}~, \hspace{15pt} \textrm{where} \hspace{15pt} {\bf Q}^{^{\textrm{PSV}}} = \left[ 
{\begin{array}{c}
U \\
V \\
W \\
X \\
\end{array}}
 \right] 
\hspace{15pt} \textrm{and} \hspace{15pt}
{\bf F}^{^{\textrm{PSV}}}= \left[ 
{\begin{array}{c}
F_{1z} \\
F_{1r} \\
F_{2z} \\
F_{2r} \\
\end{array}}
 \right] ~.
\ee
${\bf L}^{^{\textrm{PSV}}}$ is a differential operator given by
\be \label{eq212}
{\bf L}^{^{\textrm{PSV}}}={\frac {\p }{\p z}}\,\left(\,{\bf M}^{^{\textrm{PSV}}}_1\,{\frac {\p }{\p z}}+\omega p ~ {\bf M}^{^{\textrm{PSV}}}_2 \,\right) - \omega p ~ [\,{\bf M}^{^{\textrm{PSV}}}_2\,]^T \, {\frac {\p }{\p z}} + \omega^2 \, ( \, {\bf M}^{^{\textrm{PSV}}}_3 - p^2 \, {\bf M}^{^{\textrm{PSV}}}_4 \,)
\ee
where the ${\bf M}^{^{\textrm{PSV}}}_i, i=1..4$ are $4 \times 4$ matrices defined by
\begin{eqnarray} \label{eq213}
{\bf M}^{^{\textrm{PSV}}}_1 = & \left[ 
{\begin{array}{cccr}
\lambda_U + 2\,G & 0 & C & 0 \\
               0 & G & 0 & 0 \\
               C & 0 & M & 0 \\
               0 & 0 & 0 & 0
\end{array}}
 \right], ~~~ 
& {\bf M}^{^{\textrm{PSV}}}_2 = \left[ 
{\begin{array}{ccrc}
0 & -\lambda_U & 0 & -C \\
G &          0 & 0 &  0 \\
0 &         -C & 0 & -M \\
0 &          0 & 0 &  0
\end{array}}
 \right], \nonumber \\
{\bf M}^{^{\textrm{PSV}}}_3 = & \left[ 
{\begin{array}{cccc}
  \rho &      0 & \rho_f &      0 \\
     0 &   \rho &      0 & \rho_f \\
\rho_f &      0 &  \rhot &      0 \\
     0 & \rho_f &      0 & \rhot
\end{array}}
 \right], ~~~
& {\bf M}^{^{\textrm{PSV}}}_4 = \left[ 
{\begin{array}{ccrc}
G &                 0 & 0 & 0 \\
0 & \lambda_U  + 2\,G & 0 & C \\
0 &                 0 & 0 & 0 \\
0 &                 C & 0 & M
\end{array}}
 \right]. 
\end{eqnarray}
Apart from the dimensions of the matrices, we may note that equations (\ref{eq212}) and (\ref{eq213}) are very similar to the expressions obtained in the elastic case.

Formally, equation (\ref{eq211}) admits an integral solution for the displacement fields in terms of the Green's functions of the problem. 
For example, the vertical displacement $U$ at depth $z_R$ and frequency $\omega$ is given by 
\begin{eqnarray} \label{eq214}
U(z_R,\omega) & = &\int_{\mathcal{M}}      
   [~G_{1z}^{1z}(z_R,\omega;z')\,F_{1z}(z',\omega) +
     G_{1z}^{2z}(z_R,\omega;z')\,F_{2z}(z',\omega) +
\nonumber \\
& & {} \hspace{8mm}  
     G_{1z}^{1r}(z_R,\omega;z')\,F_{1r}(z',\omega) + 
     G_{1z}^{2r}(z_R,\omega;z')\,F_{2r}(z',\omega)~] ~dz'~, 
\end{eqnarray}
where $G_{ij}^{kl}(z_R,\omega,z_S)$ is the Green's function corresponding to the displacement at depth $z_R$ of phase $i$ (the values $i=1,2$ correspond to solid and relative fluid-to-solid motions, respectively) in direction $j$ ($z$ or $r$) generated by a harmonic point force $F_{kl}(z_S,\omega)$ ($k=1,2$) at depth $z_S$ in direction $l$ ($z$ or $r$). 
A total of 16 different Green's functions are needed to express the 4 displacements $U$, $V$, $W$ and $X$ in the $P-SV$-wave system (4 displacements $\times$ 4 forces). 

The integrals of equation (\ref{eq214}) are taken over the depths z' of a region $\mathcal{M}$ including the body forces ${\bf F_1}$ and ${\bf F_2}$.
In the case of a vertical point force at depth $z_S$, the expressions of the forces become
\be \label{eq216}
\left\{ \begin{array} {l}
F_{1z}(z_S,\omega) = \delta(z-z_S)~S_1(\omega) \\
F_{1r}(z_S,\omega) = 0 
\end{array} \right. 
\qquad \textrm{and} \qquad 
\left\{ \begin{array} {l}
 F_{2z}(z_S,\omega)=\delta(z-z_S)~S_2(\omega)  \\
 F_{2r}(z_S,\omega)=0 ~,
\end{array} \right.
\ee
where $S_1(\omega)$ and $S_2(\omega)$ are the Fourier transforms of the source time functions associated with forces $F_1$ and $F_2$. 
Assuming that the amplitudes of both forces are similar ($\|{\bf F_1}\| \simeq \|{\bf F_2}\|$) \citep{garambois02}, we take $S(\omega)= S_1(\omega)= S_2(\omega)$.
The displacement fields for a vertical point force can then be written in simple forms with the Green's functions:
\be \label{eq217}
\begin{array} {rcl}
 U(z_R,\omega;z_S)&=&[~ G_{1z}^{1z}(z_R,\omega;z_S) ~+~ G_{1z}^{2z}(z_R,\omega;z_S)~]~ S(\omega) \\ 
 V(z_R,\omega;z_S)&=&[~ G_{1r}^{1z}(z_R,\omega;z_S) ~+~ G_{1r}^{2z}(z_R,\omega;z_S)~]~ S(\omega) \\ 
 W(z_R,\omega;z_S)&=&[~ G_{2z}^{1z}(z_R,\omega;z_S) ~+~ G_{2z}^{2z}(z_R,\omega;z_S)~]~ S(\omega) \\ 
 X(z_R,\omega,z_S)&=&[~ G_{2r}^{1z}(z_R,\omega,z_S) ~+~ G_{2r}^{2z}(z_R,\omega,z_S)~]~ S(\omega) ~.
\end{array}
\ee
The displacement fields corresponding to a horizontal point force and to an explosive point source are similarly defined. Explosions would be represented by Green's functions $G_{ij}^{kE}$ ($i=1,2$; $j=z,r$; $k=1,2$) representing the radiation of an explosive point source $E$.

The $SH$ case is treated in exactly the same way as the $P-SV$ case. 
The corresponding second-order differential equations of motion are  
\be \label{eq2171}
\left\{ \begin{array}{l}
F_{1t} = \di{{\frac {\p }{\p z}}}\,\left[~G \,
\di{\frac {\p T}{\p z}}~\right] + \omega 
^{2} \left[~ - p^{2}\,G\,T + \rho\,T
 + \rho_f\,Y~\right]\\
F_{2t} = \omega ^{2} \left[~\rho_f\,T + \rhot\,Y~\right] , 
\end{array} \right.
\ee 
where $T=T(z_R,\omega;z_S)$ and $Y=Y(z_R,\omega;z_S)$ stand for the transverse solid and fluid-to-solid displacements; $F_{1t}(z_S,\omega)$ and $F_{2t}(z_S,\omega)$ are the transverse components of body forces ${\bf F_1}$ and ${\bf F_2}$.

As before, we can rewrite equations (\ref{eq2171}) in matrix form as 
\be \label{eq2172}
{\bf L}^{^{\textrm{SH}}} {\bf Q}^{^{\textrm{SH}}} = {\bf F}^{^{\textrm{SH}}}~, \hspace{15pt} \textrm{where} \hspace{15pt} {\bf Q}^{^{\textrm{SH}}} = \left[ 
{\begin{array}{c}
T \\
Y \\
\end{array}}
 \right] 
\hspace{15pt} \textrm{and} \hspace{15pt}
{\bf F}^{^{\textrm{SH}}}= \left[ 
{\begin{array}{c}
F_{1t} \\
F_{2t} \\
\end{array}}
 \right] ~.
\ee
Here, ${\bf L}^{^{\textrm{SH}}}$ is a linear operator defined as
\be \label{eq2173}
{\bf L}^{^{\textrm{SH}}}={\frac {\p }{\p z}}\,\left(\,{\bf M}^{^{\textrm{SH}}}_1\,{\frac {\p }{\p z}}\,\right) \,+ \omega^2 \, ( \, {\bf M}^{^{\textrm{SH}}}_2 - p^2 \, {\bf M}^{^{\textrm{SH}}}_1 \,)
\ee
where the ${\bf M}^{^{\textrm{SH}}}_i, i=1..4$, are $2 \times 2$ matrices defined by
\begin{eqnarray} \label{eq2174}
{\bf M}^{^{\textrm{SH}}}_1 = & \left[ 
{\begin{array}{cc}
G & 0 \\
0 & 0 \\
\end{array}}
\right],~ 
& {\bf M}^{^{\textrm{SH}}}_2  = \left[ 
{\begin{array}{cc}
\rho & \rho_f \\
\rho_f & \rhot \\
\end{array}}
\right].
\end{eqnarray}

\section{Fr\'echet derivatives of the plane wave reflectivity}

\subsection{Statement of the problem}

The Fr\'echet derivatives are usually introduced by considering the forward problem of the wave propagation, in which a set of synthetic seismograms ${\bf d}$ is computed for an earth model ${\bf m}$ using the non-linear relationship ${\bf d=f(m)}$.
\citet{tarantola84a} uses a Taylor series expansion to relate a small perturbation ${\bf \delta m}$ in the model parameters to a small perturbation ${\bf \delta f}$ in the wave field
\be \label{eq301}
{\bf f(m+\delta m)= f(m)+ D~\delta m +o}\,(\,\|\,{\bf \delta m}\,\|^2\,) \hspace{5mm} \textrm{or}
\hspace{5mm} {\bf \delta f = D~\delta m}
\ee
where $\bf D=\p {\bf f}\,/\,\p {\bf m}$ is the matrix of Fr\'echet derivatives.

Our aim is to compute the various Fr\'echet derivatives corresponding to slight modifications of the model parameters at a given depth. 
Considering for instance the density at depth $z$, this problem reduces, in the $P-SV$ case, to finding analytical expressions for the quantities
\be \label{eq302}
\begin{array}{l} 
\mathcal{A}^{^{\textrm{PSV}}}_1(z_R,\omega;z_S|z)=\di{\frac{\p U(z_R,\omega;z_S)}{\p \rho(z)}}\\
\mathcal{A}^{^{\textrm{PSV}}}_2(z_R,\omega;z_S|z)=\di{\frac{\p V(z_R,\omega;z_S)}{\p \rho(z)}}\\
\mathcal{A}^{^{\textrm{PSV}}}_3(z_R,\omega;z_S|z)=\di{\frac{\p W(z_R,\omega;z_S)}{\p \rho(z)}}\\
\mathcal{A}^{^{\textrm{PSV}}}_4(z_R,\omega;z_S|z)=\di{\frac{\p X(z_R,\omega;z_S)}{\p \rho(z)}} ~.
\end{array}
\ee

We can similarly define the Fr\'echet derivatives $\mathcal{B}^{^{\textrm{PSV}}}_i$, $\mathcal{C}^{^{\textrm{PSV}}}_i$, $\mathcal{D}^{^{\textrm{PSV}}}_i$, $\mathcal{E}^{^{\textrm{PSV}}}_i$, $\mathcal{F}^{^{\textrm{PSV}}}_i$ and $\mathcal{G}^{^{\textrm{PSV}}}_i$, $i=1..4$, for model parameters $\rho_f$, $\tilde{\rho}$, $C$, $M$, $\lambda_U$ and $G$. 
This is the natural choice of parameters to carry out a perturbation analysis because of the
linear dependence of these parameters with the wave equations.
We also introduce the set of Fr\'echet derivatives $\mathcal{\hat A}^{^{\textrm{PSV}}}_i$, $\mathcal{\hat B}^{^{\textrm{PSV}}}_i$, $\mathcal{\hat C}^{^{\textrm{PSV}}}_i$, $\mathcal{\hat D}^{^{\textrm{PSV}}}_i$, $\mathcal{\hat E}^{^{\textrm{PSV}}}_i$, $\mathcal{\hat F}^{^{\textrm{PSV}}}_i$, $\mathcal{\hat G}^{^{\textrm{PSV}}}_i$ and $\mathcal{\hat H}^{^{\textrm{PSV}}}_i$, $i=1..4$, corresponding to model parameters $\rho_s$, $\rho_f$, $k_0$, $\phi$, $K_s$, $K_f$, $G_s$ and $c_s$ which we will use in a second stage, and which are more convenient to use as physical parameters of the problem.
The sensitivity operators are derived by following the procedure presented in \citet{dietrich90} for the elastic case.

\subsection{Perturbation analysis}

We first present, with some detail, the perturbation analysis for the $P-SV$ case before addressing the simpler $SH$ case.
We consider small changes in the model parameters at a given depth $z$ that result in small perturbations ${\bf \Delta Q}^{^{\textrm{PSV}}} = [\,\delta U$, $\delta V$, $\delta W$, $\delta X\,]^T$ of the seismic wave field and in a modified seismic response ${\bf Q\,'\,}^{^{\textrm{PSV}}} = [\,U'$, $V'$, $W'$, $X'\,]^T$.
The latter can be written as
\be
{\bf Q\,'\,}^{^{\textrm{PSV}}} = {\bf Q}^{^{\textrm{PSV}}} + {\bf \Delta Q}^{^{\textrm{PSV}}} 
\ee
by assuming that the magnitudes of the scattered waves ${\bf \Delta Q}^{^{\textrm{PSV}}}$ are much smaller than those of the primary waves ${\bf Q}^{^{\textrm{PSV}}}$.
Considering for instance the first component of the above vector equation 
\be \label{eq303}
U'(z_R,\omega;z_S) = U(z_R,\omega;z_S) + \delta U(z_R,\omega;z_S)~, 
\ee
we can write the scattered displacement $\delta U$ as
\begin{eqnarray} \label{eq304}
\delta U(z_R,\omega;z_S) = \int_{\mathcal{M}}  &[~ & \mathcal{A}^{^{\textrm{PSV}}}_1(z_R,\omega;z_S|z) \,{\delta\rho(z)}  ~+  \mathcal{B}^{^{\textrm{PSV}}}_1(z_R,\omega;z_S|z)\,{\delta\rho_f(z)} ~+ %-{}
\nonumber \\
 & & {}  \mathcal{C}^{^{\textrm{PSV}}}_1(z_R,\omega;z_S|z) \,{\delta\rhot(z)} ~~+ \mathcal{D}^{^{\textrm{PSV}}}_1(z_R,\omega;z_S|z) \,{\delta C(z)} ~+
\nonumber \\
 & & {} \mathcal{E}^{^{\textrm{PSV}}}_1(z_R,\omega;z_S|z) \,{\delta M(z)} +  \mathcal{F}^{^{\textrm{PSV}}}_1(z_R,\omega;z_S|z) \,{\delta \lambda_U(z)} +~
\nonumber \\
 & & {} \mathcal{G}^{^{\textrm{PSV}}}_1(z_R,\omega;z_S|z) \,{\delta G(z)} ~]~dz ~.
\end{eqnarray}
With the model parametrization adopted (i.e., a linear dependence of the model parameters with the wave equation), the perturbation analysis can mainly be done in symbolic form.
Indeed, the wave operator ${\bf L}^{^{\textrm{PSV}}}$ in the perturbed medium can be written as 
\be
{\bf L\,'\,}^{^{\textrm{PSV}}} = {\bf L}^{^{\textrm{PSV}}} + {\bf \Delta L}^{^{\textrm{PSV}}}~,
\ee
so that equation (\ref{eq211}) becomes \citep{hudson81},
\be
\left( {\bf L}^{^{\textrm{PSV}}} + {\bf \Delta L}^{^{\textrm{PSV}}} \right) \left( {\bf Q}^{^{\textrm{PSV}}} + {\bf \Delta Q}^{^{\textrm{PSV}}} \right) = {\bf F}^{^{\textrm{PSV}}} ~.
\ee
We then use the single scattering (or Born) approximation to solve the above equation for ${\bf \Delta Q}^{^{\textrm{PSV}}}$ under the assumption (already used above) that $\|{\bf \Delta Q}^{^{\textrm{PSV}}}\| \ll \|{\bf Q}^{^{\textrm{PSV}}}\|$. 
Combining this approximation with equation (\ref{eq211}), we obtain
\be \label{eq305}
{\bf L}^{^{\textrm{PSV}}} \, {\bf \Delta Q}^{^{\textrm{PSV}}} \simeq -{\bf \Delta L}^{^{\textrm{PSV}}} \, {\bf Q}^{^{\textrm{PSV}}} \equiv {\bf \Delta F}^{^{\textrm{PSV}}}
\ee
This matrix equation shows that the scattered waves ${\bf \Delta Q}^{^{\textrm{PSV}}}$ due to perturbations of the material properties can be interpreted as a wave field generated by secondary body forces ${\bf \Delta F}^{^{\textrm{PSV}}}$ defined by the interaction of the incident waves with the heterogeneities (term ${\bf \Delta L}^{^{\textrm{PSV}}} \, {\bf Q}^{^{\textrm{PSV}}}$).
Moreover, this wave field propagates in the unperturbed medium represented by the wave operator ${\bf L}^{^{\textrm{PSV}}}$.
Consequently, equation (\ref{eq305}) has for each of its components a solution similar to equation (\ref{eq214}), by substituting ${\bf \Delta F}^{^{\textrm{PSV}}}$ for ${\bf F}^{^{\textrm{PSV}}}$ and ${\bf \Delta Q}^{^{\textrm{PSV}}}$ for ${\bf Q}^{^{\textrm{PSV}}}$. 
Considering again the scattered displacement $\delta U$, we have
\be \label{eq3051}
\begin{array}{lll} 
\delta U(z_R,\omega;z_S)  = & \di{\int_{\mathcal{M}} [~} &
  G_{1z}^{1z}(z_R,\omega;z')\,\delta F_{1z}(z',\omega) +
     G_{1z}^{2z}(z_R,\omega;z')\,\delta F_{2z}(z',\omega) + \\
& &    
     G_{1z}^{1r}(z_R,\omega;z')\,\delta F_{1r}(z',\omega) +
     G_{1z}^{2r}(z_R,\omega;z')\,\delta F_{2r}(z',\omega)~] ~dz'~, 
\end{array}
\ee
where the secondary Born sources $\delta F_{1z}, \delta F_{1r},\delta F_{2z}, \delta F_{2r}$ are obtained from equation (\ref{eq210}):
\be \label{eq306}
{\bf \Delta F}^{^{\textrm{PSV}}} = \left( \begin{array}{c} 
{\delta F_{1z}} \\
{\delta F_{1r}} \\
{\delta F_{2z}} \\
{\delta F_{2r}}
\end{array} \right)
= \left( \begin{array}{lll}
-\di{\frac {\p }{\p z}}\,\left[~({\delta \lambda_U} + 
   2 \,{\delta G} )\,\di{\frac {\p U}{\p z}} + {\delta C}\,\di{\frac {\p W}{\p z}} - 
   \omega p\,({\delta \lambda_U} \,V + {\delta  C}\,X)~\right] \\
   \vspace{6pt}
   \lefteqn ~~~~~ + \, \omega p\,{\delta G} \,\di{\frac {\p V}{\p z}} - 
   \omega ^2 \left[~{\delta \rho}\,U - p^2\,{\delta G} \,U + 
   {\delta \rho_f}\,W~\right]\\
-\di{\frac {\p }{\p z}}\,\left[~{\delta G} \,
   \di{\frac {\p V}{\p z}} + \omega p\,{\delta G} \,U~\right] - 
   \omega p\,\left[~{\delta \lambda_U} \,\di{\frac {\p U}{\p z}} + 
   {\delta C}\, \di{\frac {\p W}{\p z}}~\right]\\
   \vspace{6pt}
   \lefteqn ~~~~~ - \, \omega ^2\,\left[~{\delta \rho}\,V - 
   p^2 \,({\delta \lambda_U} + 2\,{\delta G} )\,V - 
   p^2\,{\delta C}\,X~\right]\\
-\di{\frac {\p }{\p z}}\,\left[~{\delta C}\,
   \left(\,\di{\frac {\p U}{\p z}} - \omega p\,V \,\right) + {\delta M}\,
   \left(\,\di{\frac {\p W}{\p z}} - \omega p\,X \,\right) ~\right]\\
   \vspace{6pt}
   \lefteqn ~~~~~ - \, \omega ^2\,\left[~{\delta \rho_f}\,U + 
   {\delta \tilde{\rho}}\,W~\right]\\
-\omega p\,\left[~{\delta C}\,\di{\frac {\p U}{\p z}} + 
   {\delta M}\,\di{\frac {\p W}{\p z}}~\right] \\ 
   \vspace{6pt}
   \lefteqn ~~~~~ - \, \omega^2 \, \left[~{\delta \rho_f}\,V + 
   {\delta \tilde{\rho}}\,X - p^2\,({\delta C}\,V + {\delta M}\,X)~\right]
\end{array} \right).
\ee
By inserting these expressions into equation (\ref{eq3051}) and integrating by parts in order to separate the contributions in $\delta \rho$, $\delta \rho_f$, $\delta \tilde{\rho}$, $\delta C$, $\delta M$, $\delta \lambda_U$ and $\delta G$, we obtain an integral representation of the scattered wave field $\delta U$ that we can directly identify with equation (\ref{eq304}) to get the Fr\'echet derivatives $\mathcal{A}^{^{\textrm{PSV}}}_1$, $\mathcal{B}^{^{\textrm{PSV}}}_1$, $\mathcal{C}^{^{\textrm{PSV}}}_1$, $\mathcal{D}^{^{\textrm{PSV}}}_1$, $\mathcal{E}^{^{\textrm{PSV}}}_1$, $\mathcal{F}^{^{\textrm{PSV}}}_1$ and
$\mathcal{G}^{^{\textrm{PSV}}}_1$ corresponding to displacement $U$:
\be \label{eq307}
\begin{array}{lll}
\mathcal{A}^{^{\textrm{PSV}}}_1 &=& - \omega^2\,[~U\,{G_{1z}^{1z}} + V\,{G_{1z}^{1r}}~]\\ %rho
\vspace{3pt}
\mathcal{B}^{^{\textrm{PSV}}}_1 &=& - \omega^2\,[~W \,{G_{1z}^{1z}} +  
   U\,{G_{1z}^{2z}} + V\,{G_{1z}^{2r}}~]\\   %rhof
\vspace{3pt}
\mathcal{C}^{^{\textrm{PSV}}}_1 &=& - \omega ^2\,[~W\,{G_{1z}^{2z}} + 
   X\,G_{1z}^{2r}~]\\ %\tilde{\rho}
\vspace{3pt}
\mathcal{D}^{^{\textrm{PSV}}}_1 &=& \left[~\di{\frac {\p W}{\p z}} -\omega p\,X
   ~\right]\,\left[~\di{\frac {\p {G_{1z}^{1z}}}{\p z}}-\omega p\,{G_{1z}^{1r}}~\right] +
   \left[~\di{\frac {\p U}{\p z}}-\omega p\,V ~\right]\,
   \left[~\di{\frac {\p {G_{1z}^{2z}}}{\p z}}- \omega p\,{G_{1z}^{2r}}~\right]\\ %C
\vspace{3pt}
\mathcal{E}^{^{\textrm{PSV}}}_1 &=& \left[~\di{\frac {\p  W}{\p z}} - 
   \omega p\,X~\right]\,\left[~\di{\frac {\p {G_{1z}^{2z}}}{\p z}} -
   \omega p\,{G_{1z}^{2r}}~\right]\\     %M
\vspace{3pt}
\mathcal{F}^{^{\textrm{PSV}}}_1 &=& \left[~\di{\frac {\p U}{\p z}} - 
   \omega p\,V~\right]\,\left[~\di{\frac {\p {G_{1z}^{1z}}}{\p z}} -
   \omega p \,{G_{1z}^{1r}}~\right]\\ %lambda u
\vspace{3pt}
\mathcal{G}^{^{\textrm{PSV}}}_1 &=& \left[~\di{\frac {\p V}{\p z}} + \omega p\,U ~\right]\,
   \left[~ \di{\frac {\p {G_{1z}^{1r}}}{\p z}} + \omega p\,
   {G_{1z}^{1z}}~\right] + 2 \left[\,
   \di{\frac {\p U}{\p z}}\,\di{\frac {\p {G_{1z}^{1z}}}{\p z}} + 
   \omega^2 p^2\,V\,{G_{1z}^{1r}} \,\right]~.%G
\end{array}
\ee 
Here, $U=U(z,\omega;z_S)$, $V=V(z,\omega;z_S)$, $W=W(z,\omega;z_S)$ and $X=X(z,\omega;z_S)$ denote the incident wave fields at the model perturbations.
These wave fields can be expressed in terms of Green's functions using equations (\ref{eq217}) for a vertical point force, and similarly for a horizontal point force or for an explosive point source.
Expressions $G_{ij}^{kl}=G_{ij}^{kl}(z_R, \omega;z)$ represent the Green's functions conveying the scattered wave fields from the inhomogeneities to the receivers, as noted before. 
A total of 32 Green's functions (16 for up-going waves and another 16 for down-going waves) are to be computed to completely solve these equations.

The Fr\'echet derivatives for the radial displacement $V$ are easily deduced from expressions (\ref{eq307}) by changing $G_{1z}^{kl}$ to $G_{1r}^{kl}$. 
In the same way, the Fr\'echet derivatives corresponding to the vertical and radial relative fluid-to-solid displacements are obtained by substituting $G_{2j}^{kl}$ for $G_{1j}^{kl}$. 
As a first verification of the formulas, we note that the expressions corresponding to perturbations of parameters $\rho$, $\lambda_U$ and $G$ ($\mathcal{A}^{^{\textrm{PSV}}}_i$, $\mathcal{F}^{^{\textrm{PSV}}}_i$ and $\mathcal{G}^{^{\textrm{PSV}}}_i$, respectively) are in perfect agreement with the Fr\'echet derivatives formulas computed for the density $\rho$ and Lam\'e parameters $\lambda$ and $\mu$ in the elastic case \citep{dietrich90}.

\subsection{Fr\'echet derivatives for relevant parameters}

As mentioned above, the set of 7 parameters ($\rho$, $\rho_f$, $\tilde{\rho}$, $C$, $M$, $\lambda_U$ and $G$) used in the perturbation analysis was primarily chosen to considerably simplify the derivation of expressions (\ref{eq307}). 
In practice, and as with the Lam\'e parameters in the elastic case, it is more convenient to consider model parameters which are easier to measure or estimate. 
We introduce here a new set of 8 parameters, namely $\rho_s$, $\rho_f$, $k_0$, $\phi$, $K_f$, $K_s$, $G_s$ and $c_s$ which are more naturally related to the solid and fluid phases. 
Furthermore, these parameters are independent from each other in terms of their mechanical or hydrological meaning.
They are all real and frequency independent, contrary to $\rhot$. 
Fluid viscosity is not considered here as the Fr\'echet derivatives with respect to the fluid viscosity and to the permeability are strictly proportional. 
We concentrate on the permeability by stressing that the comments made for the permeability will also be valid for the fluid viscosity.
 
To obtain the Fr\'echet derivatives corresponding to the new set of parameters from expressions (\ref{eq307}), we construct the $8 \times 7$ Jacobian matrix $\bf{J}$ whose coefficients are formally defined by
\be \label{eq3070}
 J_{ij}=\frac{\p p_i}{\p p'_j}~,
\ee 
where $p_i, i=1..7$, stands for one of the parameters $\rho$, $\rho_f$, $\tilde{\rho}$, $C$, $M$,  $\lambda_U$ or $G$, and where $p'_j, j=1..8$, represents one of the parameters $\rho_s$, $\rho_f$, $k_0$, $\phi$, $K_s$, $K_f$, $G_s$ or $c_s$.
The elements of matrix ${\bf J}$ are determined from equations (\ref{eq202}) to (\ref{eq207}).\\
The Fr\'echet derivatives with respect to the new set of parameters for the vertical component of the solid displacement $U$ are then defined by
\be \label{eq3071} \left(
\begin{array}{c}
\mathcal{\hat A}^{^{\textrm{PSV}}}_i \\
\mathcal{\hat B}^{^{\textrm{PSV}}}_i \\
\mathcal{\hat C}^{^{\textrm{PSV}}}_i \\
\mathcal{\hat D}^{^{\textrm{PSV}}}_i \\
\mathcal{\hat E}^{^{\textrm{PSV}}}_i \\
\mathcal{\hat F}^{^{\textrm{PSV}}}_i \\
\mathcal{\hat G}^{^{\textrm{PSV}}}_i \\
\mathcal{\hat H}^{^{\textrm{PSV}}}_i
\end{array} \right)
= {\bf J} \left(
\begin{array}{c}
\mathcal{A}^{^{\textrm{PSV}}}_i \\
\mathcal{B}^{^{\textrm{PSV}}}_i \\
\mathcal{C}^{^{\textrm{PSV}}}_i \\
\mathcal{D}^{^{\textrm{PSV}}}_i \\
\mathcal{E}^{^{\textrm{PSV}}}_i \\
\mathcal{F}^{^{\textrm{PSV}}}_i \\
\mathcal{G}^{^{\textrm{PSV}}}_i
\end{array} \right)~.
\ee
The combinations involved in equation (\ref{eq3071}) significantly complicate the expressions of the Fr\'echet derivatives corresponding to the new set of parameters.
For sake of clarity and simplification, we introduce the following quantities
\begin{eqnarray} \label{eq3072}
\Omega & = & \sqrt{n_J - 4\,i\,{\frac {\omega}{\omega_c }}}~ 
           \nonumber \\
a & = & (1 + c_s \phi)\,K_s + c_s (1 - \phi)\,K_f  
      \nonumber \\
b & = & \phi ^2\,\di \left\{ 4\,(2 + 3\,c_s)\,G_s\,a^2  - 3\,(2 + 3\,c_s\,\phi)^2 \,K_s\, (1+c)\, \left[ {K_s}^2\,- c_s\,{K_f}^2\, +   
        (c_s-1)\,K_s\,K_f\,\right]\di \right\} 
       \nonumber \\
c & = & \phi ^2\,\left[\,(2 + 3\,c_s\,\phi)^2\,K_s\,({K_s} - \,{K_f})^2\,- 4\,a^2\,G_s\right]~. 
\end{eqnarray}
With these parameters, the final expressions of the Fr\'echet derivatives
$\mathcal{\hat A}^{^{\textrm{PSV}}}_1$, $\mathcal{\hat B}^{^{\textrm{PSV}}}_1$, $\mathcal{\hat C}^{^{\textrm{PSV}}}_1$, $\mathcal{\hat D}^{^{\textrm{PSV}}}_1$, $\mathcal{\hat E}^{^{\textrm{PSV}}}_1$, $\mathcal{\hat F}^{^{\textrm{PSV}}}_1$, $\mathcal{\hat G}^{^{\textrm{PSV}}}_1$ and $\mathcal{\hat H}^{^{\textrm{PSV}}}_1$ with respect to model parameters $\rho_s$, $\rho_f$, $k_0$, $\phi$, $K_s$, $K_f$, $G_s$ and $c_s$ are
\begin{eqnarray} \label{eq3073}
\mathcal{\hat A}^{^{\textrm{PSV}}}_1 & = & (1 - \phi )~\mathcal{A}^{^{\textrm{PSV}}}_1 
                                         \nonumber \\
\mathcal{\hat B}^{^{\textrm{PSV}}}_1 & = &  \phi ~\mathcal{A}^{^{\textrm{PSV}}}_1 + 
                                         \mathcal{B}^{^{\textrm{PSV}}}_1 + 
                                         {\di \frac {2 + \Omega}{\Omega}}\,F~
                                         \mathcal{C}^{^{\textrm{PSV}}}_1 \nonumber \\
\mathcal{\hat C}^{^{\textrm{PSV}}}_1 & = & -{\di \frac {\left(i\,n_J + 2\,
                                         {\di \frac {\omega }{{\omega_c}}} \right)\,\eta \,}{
                                         {k_0}^{2}\,\Omega \,\omega}}~
                                         \mathcal{C}^{^{\textrm{PSV}}}_1 \nonumber \\
\mathcal{\hat D}^{^{\textrm{PSV}}}_1 & = & \frac{1}{a^2\,\phi^2\,(2+3\,c_s\,\phi)^2}\,
                                         \left\{~a^2\,\phi^2\,(\rho_s - \rho_f)\, 
                                         (2+3\,c_s\,\phi)^2~
                                         \mathcal{A}^{^{\textrm{PSV}}}_1 -~ 2\,a^2\,\phi^2\,
                                         (2 + 3\,c_s)\,G_s~
                                         \mathcal{G}^{^{\textrm{PSV}}}_1 \right. \nonumber \\
& & {}                                   -~(2+3\,c_s\,\phi)^2\,\Big[\,(1 + c_s\,\phi )^2\,K_s +
                                         c_s\,(1 - 2\,\phi - c_s\,\phi ^2)\,K_f\, \Big]\, 
                                         	K_s\,{K_f}~\mathcal{E}^{^{\textrm{PSV}}}_1 
                                         \nonumber\\
& & {}                                   \left. -~\phi ^{2}\,c_s\,(1 + c_s)\,
                                         (2+3\,c_s\,\phi)^2\,
                                         (K_s - K_f)\,K_s\,K_f~
                                         \mathcal{D}^{^{\textrm{PSV}}}_1 +~ 
                                         \frac {1}{3}\,b~
                                         \mathcal{F}^{^{\textrm{PSV}}}_1 ~\right\} \nonumber \\
\mathcal{\hat E}^{^{\textrm{PSV}}}_1 & = & \frac{1-\phi}{a^2\,\phi} \,\left\{~\phi \, c_s \, 
                                         (1 + c_s)\,{K_f}^2~
                                         \mathcal{D}^{^{\textrm{PSV}}}_1 +~ c_s\,
                                         (1+c_s\,\phi)\,{K_f}^2~
                                         \mathcal{E}^{^{\textrm{PSV}}}_1 \right. \\
& & {}                                   \left. +~\phi\,\Big[\,(1+c_s\,\phi)\,{K_s}^2 + 
                                         c_s\,(c_s + \phi)\,{K_f}^2 + 2 \,c_s\,(1-\phi)\,
                                         {K_s}\,{K_f}\, \Big]~
                                         \mathcal{F}^{^{\textrm{PSV}}}_1~\right\}
                                         \nonumber \\
\mathcal{\hat F}^{^{\textrm{PSV}}}_1 & = & \frac{1}{a^2\,\phi}~
                                         \left\{~(1 + c_s\,\phi)\,{K_s}^2\,\Big[\,
                                         \phi\,(1 + c_s)~
                                         \mathcal{D}^{^{\textrm{PSV}}}_1 +~(1 + c_s\,\phi)~
                                         \mathcal{E}^{^{\textrm{PSV}}}_1 \Big] \right.
                                         \nonumber \\
& & {}                                   \left. +~\Big[\,\phi\,(c_s + \phi)\,(1 + c_s\,\phi)
                                         - c_s\,(1 - \phi)^2\,\Big]\,
                                         {K_s}^2~\mathcal{F}^{^{\textrm{PSV}}}_1 \right\}
                                         \nonumber \\
\mathcal{\hat G}^{^{\textrm{PSV}}}_1 & = & 2\,\frac{1 - \phi}{2 + 3\,c_s\,\phi}\,
                                         \Big[{-\di \frac {2}{3}} \,
                                         \mathcal{F}^{^{\textrm{PSV}}}_1 +~
                                         \mathcal{G}^{^{\textrm{PSV}}}_1 \Big] \nonumber \\
\mathcal{\hat H}^{^{\textrm{PSV}}}_1 & = & \frac{1-\phi}{a^2\,\phi \,(2 + 3\,c_s\,\phi)^2}\, 
                                         \left\{~(2+3\,c_s\,\phi)^2 \,K_s\,K_f\,                                                         \Big[\,\phi\,(K_s - K_f) ~
                                         \mathcal{D}^{^{\textrm{PSV}}}_1 -~ K_f~
                                         \mathcal{E}^{^{\textrm{PSV}}}_1 \,\Big]
                                         \right. \nonumber \\
& & {}                                   \left.-~c~\mathcal{F}^{^{\textrm{PSV}}}_1 -~
                                         6\,a^2\,\phi^2\,G_s~
                                         \mathcal{G}^{^{\textrm{PSV}}}_1 ~\right\} \nonumber
\end{eqnarray}

We note that the Fr\'echet derivatives $\mathcal{\hat B}^{^{\textrm{PSV}}}_1$ and $\mathcal{\hat C}^{^{\textrm{PSV}}}_1$ with respect to fluid density $\rho_f$ and permeability $k_0$ are complex due to the role of these parameters in the attenuation and dispersion of seismic waves. \\
In addition, formulas (\ref{eq3073}) can be further simplified if source and receivers are located at the same depth $z_0=z_R=z_S$. In this case, we can take advantage of the representation of the incident wave fields $U$, $V$, $W$ and $X$ in terms of the Green's functions (see equations \ref{eq217}), and use the reciprocity theorem: 
\be \label{eq215}
G_{ij}^{kl}(z_R,\omega;z_S)=G_{kl}^{ij}(z_S,\omega;z_R)~.
\ee
The number of Green's functions required to describe the wave propagation then reduces from 32 to 16.
These simplifications are straightforward and are not developed here.

\subsection{$\mathbf{SH}$ case}

We follow the same procedure as above to derive the Fr\'echet derivatives of the solid displacement $T$ and relative fluid-to-solid displacement $Y$ in the $SH$ case. 
We denote by $\mathcal{A}^{^{\textrm{SH}}}_{i}$, $\mathcal{B}^{^{\textrm{SH}}}_{i}$,  $\mathcal{C}^{^{\textrm{SH}}}_{i}$, $\mathcal{D}^{^{\textrm{SH}}}_{i}$, $\mathcal{E}^{^{\textrm{SH}}}_{i}$, $\mathcal{F}^{^{\textrm{SH}}}_{i}$ and $\mathcal{G}^{^{\textrm{SH}}}_{i}$, respectively, the Fr\'echet derivatives with respect to model parameters $\rho$, $\rho_f$, $\tilde{\rho}$, $M$, $K_U$, $G$ and $C$, where subscript $i=1$ refers to the solid displacement $T$ and subscript $i=2$ refers to the relative fluid-to-solid displacement $Y$. 
We also introduce the notations $\mathcal{\hat A}^{^{\textrm{SH}}}_{i}$, $\mathcal{\hat B}^{^{\textrm{SH}}}_{i}$,  $\mathcal{\hat C}^{^{\textrm{SH}}}_{i}$, $\mathcal{\hat D}^{^{\textrm{SH}}}_{i}$, $\mathcal{\hat E}^{^{\textrm{SH}}}_{i}$, $\mathcal{\hat F}^{^{\textrm{SH}}}_{i}$, $\mathcal{\hat G}^{^{\textrm{SH}}}_{i}$ and $\mathcal{\hat H}^{^{\textrm{SH}}}_{i}$ for the Fr\'echet derivatives relative to our alternative set of model parameters $\rho_s$, $\rho_f$, $k_0$, $\phi$, $K_s$, $K_f$, $G_s$ and $c_s$.
The perturbation analysis leads to the following expressions of the Fr\'echet derivatives of the transverse solid displacement $T$:
\begin{eqnarray}
\begin{array}{l}
\mathcal{A}^{^{\textrm{SH}}}_{i}=-\omega ^{2}\,T\,G_{1t}^{1t}\\
\mathcal{B}^{^{\textrm{SH}}}_{i}=-\omega^2\, [\,Y\,G_{1t}^{1t} \,+\,T\,G_{1t}^{2t} ]\\
\mathcal{C}^{^{\textrm{SH}}}_{i}=-\omega^2 \,Y\, G_{1t}^{2t} \\
\mathcal{D}^{^{\textrm{SH}}}_{i}=0\\
\mathcal{E}^{^{\textrm{SH}}}_{i}=0\\
\mathcal{F}^{^{\textrm{SH}}}_{i}=0\\
\mathcal{G}^{^{\textrm{SH}}}_{i}= \di{{\frac {\p T}{\p z}}}\,\di{{\frac {\p 
G_{1t}^{1t}}{\p z}}\,} + \omega ^{2}\,p^{2
}\,T\,G_{1t}^{1t}\
\end{array} 
\end{eqnarray}
We note, as before, that the displacements $T$ and $Y$ can be expressed in terms of the Green's functions $G_{it}^{jt}$, where subscripts $i=1$ and $j=1$ relate to solid displacement and force, whereas subscripts $i=2$ and $j=2$ relate to relative fluid-to-solid displacement and force. 
Subscript $t$ stands for the tangential displacement or force.
The transformation of these expressions with the Jacobian matrix ${\bf J}$ defined in equation (\ref{eq3070}) finally yields:
\begin{eqnarray}
\begin{array}{l}
\mathcal{\hat A}^{^{\textrm{SH}}}_{i}= (1 - \phi )\,A_{sh1} \\
\mathcal{\hat B}^{^{\textrm{SH}}}_{i}= \phi \,A_{sh1} +B_{sh1} + {\di \frac {2 + \Omega}\Omega}\,F\,C_{sh1} \\
\mathcal{\hat C}^{^{\textrm{SH}}}_{i}= -\di{\frac{\di{(i n_J+2 \frac{\omega}{w_c})\,\eta}}{k_0^2 \,\Omega\,\omega}}\,C_{sh1}\\
\mathcal{\hat D}^{^{\textrm{SH}}}_{i}= (\rho_s-\rho_f)\,A_{sh1}-\,2 \,G_s\di{\frac{2+3\, c_s}{(2+3 \,c_s\, \phi)^2}}\,G_{sh1}  \\
\mathcal{\hat E}^{^{\textrm{SH}}}_{i}=0\\
\mathcal{\hat F}^{^{\textrm{SH}}}_{i}=0\\
\mathcal{\hat G}^{^{\textrm{SH}}}_{i}=2 \,\di{\frac{1-\phi}{2+3 \,c_s \,\phi}}\,G_{sh1}\\
\mathcal{\hat H}^{^{\textrm{SH}}}_{i}=-6\,G_s\,\phi\di{\frac{1-\phi}{(2+3 \,c_s\, \phi)^2}}\,G_{sh1}\\
\end{array} 
\end{eqnarray}
We note that the Fr\'echet derivatives with respect to $K_s$ and $K_f$ are zero since these parameters have no influence on shear waves.

\section{Numerical simulations and accuracy tests}

\subsection{Fr\'echet derivatives {\em vs} discrete perturbations}

As mentioned in the introduction, the Green's functions for layered media are computed with the Generalized Reflection and Transmission Matrix Method of \citet{kennett79} which yields the plane-wave response in the frequency--ray parameter (or horizontal wavenumber) domain. 
In our numerical applications, the Fr\'echet derivatives are first calculated in the frequency--wavenumber domain before being transformed into the time--distance domain with the discrete wavenumber integration method \citep{bouchon81}. 

In order to test our analytical formulations and assess their limitations, we compare the differential seismograms computed with the first-order Fr\'echet derivative approach with the seismograms obtained by introducing discrete perturbations in the medium properties.
Thus, considering for instance the vertical component of the solid displacement in the $P-SV$ case, the partial derivative $\p U/\p p_j$ with respect to parameter $p_j$ ($j=1..8$, according to the parameter set considered) can be approximated by the following finite difference expression: 
\begin{eqnarray} \label{eq401}
 \frac{\p U(z_R,\omega;z_S)}{\p p_j}
 & & {}\simeq  \frac{U^{p_j+\Delta p_j}(z_R,\omega;z_S)-U^{p_j}(z_R,\omega;z_S)}{\Delta p_j} ~,
\end{eqnarray}
where $\Delta p_j$ represents a small perturbation of parameter $p_j$.
 
The similarity between the seismograms computed with the two approaches indicated in the left- and right-hand sides of equation (\ref{eq401}) is evaluated from the correlation coefficients between the traces.

\subsection{Uniform medium}

We first consider the simple case of small perturbations $\delta \rho_s$, $\delta \rho_f$, $\delta k_0$, $\delta \phi$, $\delta K_s$, $\delta K_f$, $\delta G_s$ and $\delta c_s$ within a thin slab embedded in an infinite uniform medium. 
The slab thickness is of the order of one twentieth of the dominant wavelength of the $P$-waves (i.e., 1 m). 
The amplitude of the relative perturbations $\Delta p_j / p_j$ is 10 \% for each of the parameters considered. 
Source and receivers are located at the same depth, 50 m above the model perturbation. 
The parameters of the uniform model are listed in table 1.
The seismic response of the thin slab is considered as the reference for the comparisons with the Fr\'echet derivative seismograms.

\begin{table}
\caption{Model parameters of the infinite medium used for the numerical tests of the Fr\'echet derivatives.}
\begin{tabular}{cccccccc} 
\hline
$\phi \,(~)$ & $k_0 \,(m^2)$ & $\rho_f \,(kg/m^3$) & $\rho_s \,(kg/m^3)$ & $K_s \,(GPa)$ &  $G_s \,(GPa)$  & $K_f\, (GPa)$   &  $c_s \,(~)$  \\
 $0.20$ & $10^{-12}$ & $1000$ & $2700$ & $35$ &  $25$ &  $2.2$ & $50$  \\ 
\hline
\end{tabular} 

\end{table} 

\begin{figure}[!hbp]
\centering
\includegraphics[height=10cm]{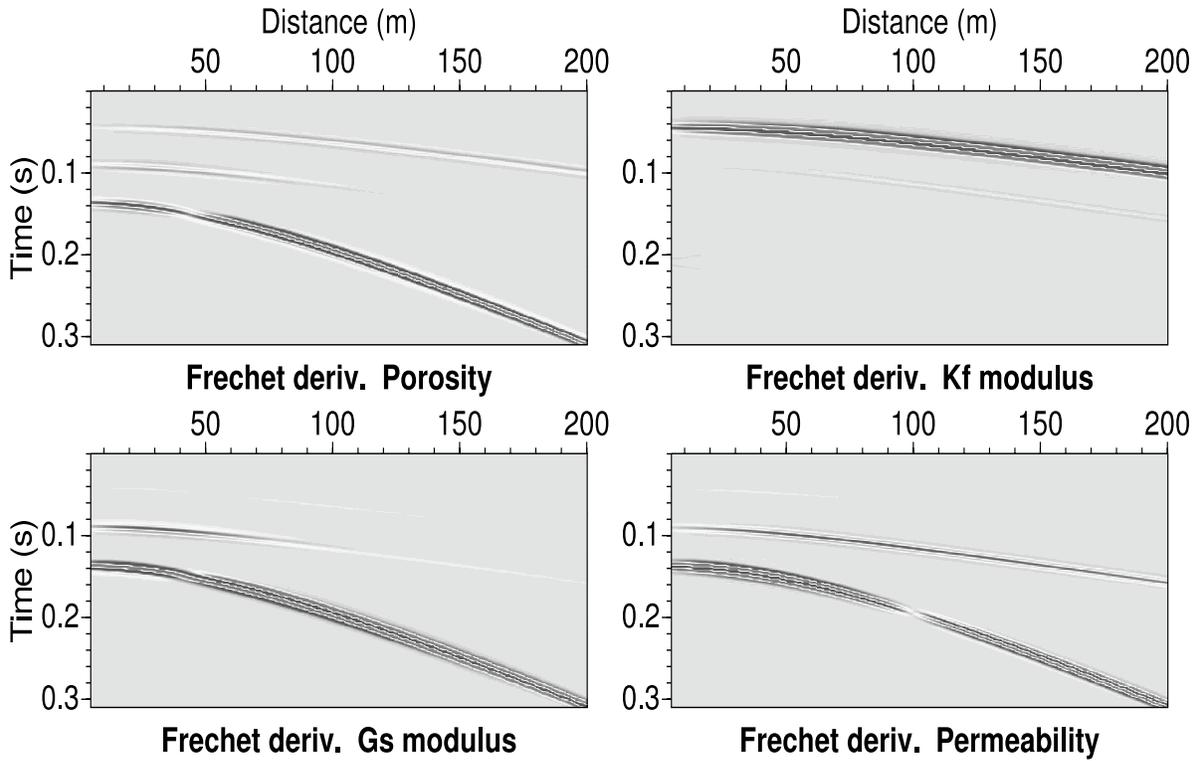}
\caption{Seismic sections corresponding to the first-order perturbation of the vertical displacement with respect to porosity $\phi$, fluid modulus $K_f$, mineral shear modulus $G_s$ and permeability $k_0$ in the uniform and infinite medium described in table 1. The seismic excitation is a vertical point force.}
\label{fig1}
\end{figure}

The simulations shown in figure \ref{fig1} include $P_{fast}$-, $P_{slow}$- and $S$-waves whose computed velocities are respectively equal to 2250, 130 and 750  m/s at a frequency of 85 Hz.
The slow $P$-waves are not visible, but three reflected waves (compressional $PP$, converted $PS$ and $SP$, and shear $SS$) are easily identified in the four sections displayed in figure \ref{fig1}. 
It is seen that a small perturbation of the fluid modulus $K_f$ has no influence on shear waves.
The same behavior is observed for the solid modulus $K_s$ (not shown).
On the contrary, slight changes in the other parameters mainly generate $SS$ reflections, as noted in particular for the porosity $\phi$, mineral shear modulus $G_s$ and permeability $k_0$ in figure \ref{fig1}. 
We also observed that the differential seismograms are very similar for the following pairs of perturbations: {\em i)} consolidation parameter and porosity; {\em ii)} fluid and solid moduli; {\em iii)} fluid and solid densities.
In addition, we found that the correlation coefficients between the Fr\'echet derivative and discrete perturbation seismograms are greater than 99 \% for most model parameters at all source-receiver offsets. 
The only exception concerns the Fr\'echet derivative with respect to the permeability which shows correlation coefficients between 60 and 95 \% in the  $P-SV$ case, depending on the source-receiver offset. 
However, this operator appears more accurate in the $SH$ case. 
By and large, the tests performed in a uniform medium validate our analytical expressions derived in sections III.C and III.D.

We further check the accuracy and stability of the first-order sensitivity operators by modifying the amplitude of the discrete perturbations, with the following results:  {\em i)} the Fr\'echet derivatives with respect to parameters that only influence $P$-waves are more accurate than the Fr\'echet derivatives with respect to parameters that influence both $P$- and $S$-waves. 
{\em ii)} Strong perturbations of the solid and fluid moduli $K_s$ and $K_f$ do not produce any distortion of the waveforms, but merely result in a global increase of the amplitudes of the discrete perturbation seismograms. 
{\em iii)} For strong amplitude perturbations, the Fr\'echet derivatives are more stable at near offsets (i.e., for small angles of incidence) than at large offsets. 
We interpret this observation as being due to the nonlinearity inherent to large offsets where the wave fields interact more strongly with the subsurface structure.
However, as an exception to this rule, the Fr\'echet derivatives with respect to permeability appear more stable at large source-receiver offsets. 
This may be explained by the fact that a perturbation of the permeability mainly influences the wave attenuation and dispersion and therefore has a stronger effect for longer travel paths. The deterioration of the seismograms correlation with decreasing $k_0$ (and corresponding decrease in wave attenuation), is another indication of the weak influence of the permeability on the seismograms.
{\em iv)} For strong perturbations, the accuracy of the Fr\'echet derivatives deteriorates for specific offsets corresponding to critical angles. 
This reduced accuracy manifests itself by modifications of the relative amplitudes of $P-$ and $S-$waves rather than by waveform changes. 
{\em v)} When checked against discrete perturbations of positive and negative amplitude of the same magnitude, the first-order approximations do not show exactly the same accuracy. 
In general, the Fr\'echet derivative seismograms obtained for positive perturbations display a better accuracy.
{\em vi)} As a general rule, the first-order approximations appear remarkably accurate for amplitude perturbations up to 20 \% in absolute value.

We now consider the robustness of the Fr\'echet derivative seismograms with respect to the thickness of the perturbed layer.
In our uniform model, the wavelengths $\lambda_P$ and $\lambda_S$ corresponding to $P$- and $S$-waves are respectively equal to 26 m and 9 m at the dominant frequency of the Ricker wavelet used in the simulation. 
Our computations show that the Fr\'echet derivative seismograms are very well correlated with the discrete perturbation seismograms until the thickness of the perturbed layer reaches about 20\% of the dominant wavelength $\lambda_P$ (that is, $\simeq$ 5 m) for parameters $K_s$ and $K_f$, and 20\% of the dominant wavelength $\lambda_S$ (that is, $\simeq$ 2 m) for all other parameters that influence both $P$- and $S$-waves.
Thus, we observe that the Fr\'echet derivatives with respect to $K_s$ and $K_f$ are more robust than the other expressions with respect to departures from the "small and localized perturbation" assumption of the Born approximation.

\subsection{Complex model}

\begin{figure}[htp]
\centering
\includegraphics[height=7cm]{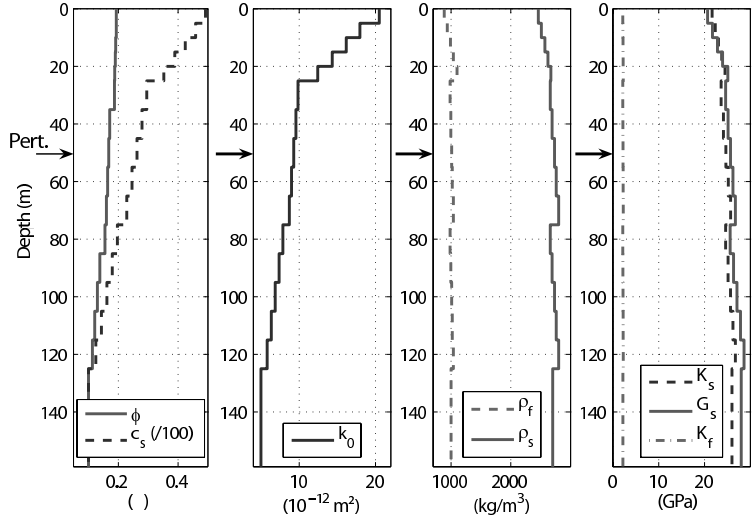}
\vspace{-0.5cm}\caption{16-layer model used for the numerical tests of the Fr\'echet derivative formulation in a stratified medium. The figure shows the distributions of the eight model parameters as a function of depth. }
\label{fig2}
\end{figure}

\begin{figure}[!hbp]
\centering
\includegraphics[height=15cm]{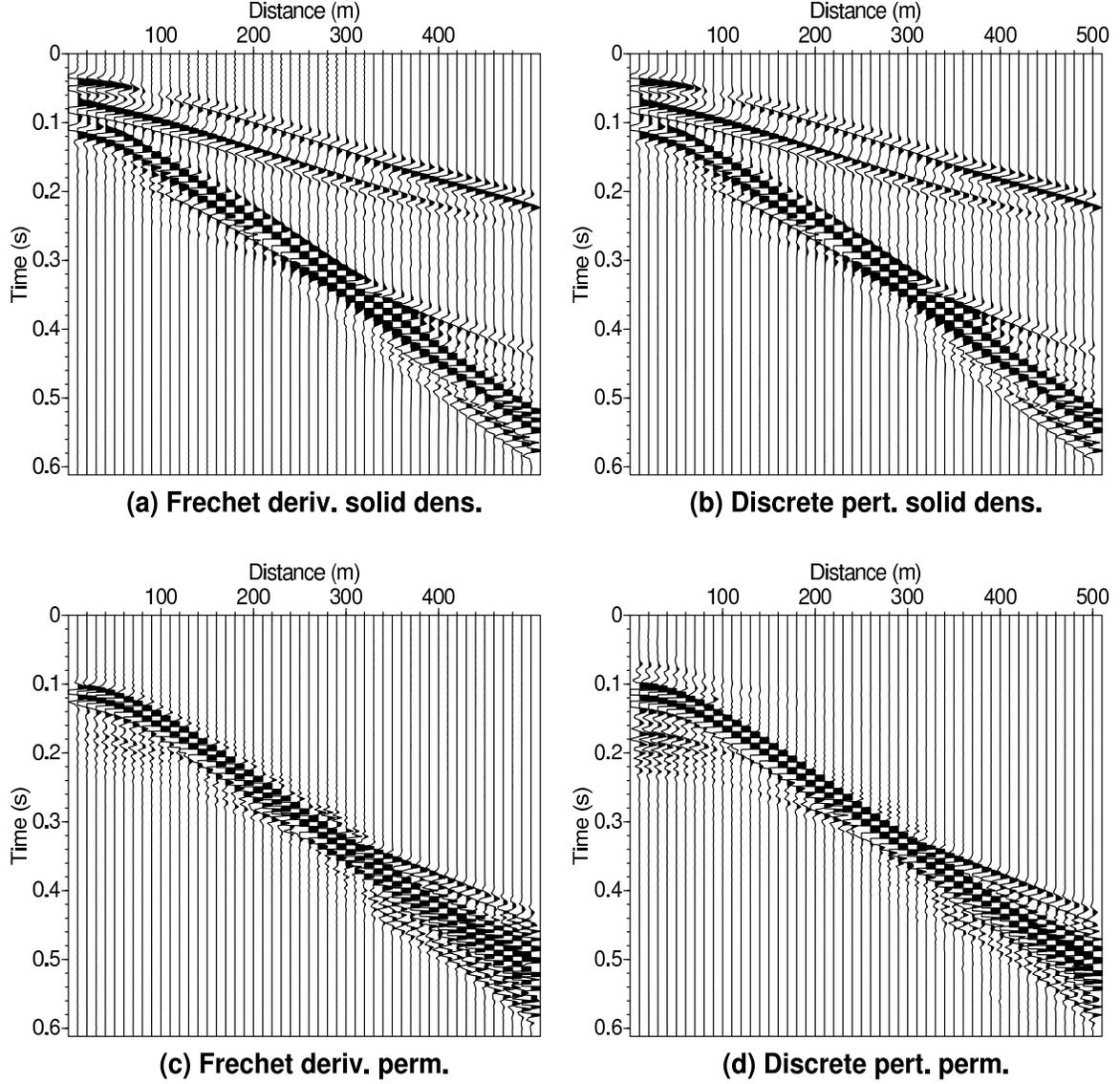} 
\caption{{\em (a)} and {\em (b)} Seismic sections obtained with the Fr\'echet derivatives and with the discrete perturbation methods for a perturbation of the solid density $\rho_s$ at $z=50$ m in the 16-layer model depicted in figure \ref{fig2}. The seismograms represent the vertical displacement generated by a vertical force ($P-SV$ wave system). {\em (c)} and {\em (d)} Same for a perturbation of the permeability $k_0$. In this case, the seismograms represent the horizontal transverse displacement generated by a horizontal transverse force ($SH$ wave system).}
\label{fig3}
\end{figure}

We used the 16-layer model shown in figure \ref{fig2} to numerically check the stability and accuracy of the Fr\'echet derivative formulas in a more complex structure. 
In this model, the perturbed layer is at a depth of 50 m, source and receivers being located near the surface. 
Figure \ref{fig3} presents the seismic sections obtained with the Fr\'echet derivative and discrete perturbation methods for slight changes of the solid density $\rho_s$ and permeability $k_0$. 
On the whole, the comparison of the waveforms obtained with both methods is very satisfactory in spite of some differences observed at small offsets for the perturbation of the permeability. 

\begin{figure}[!hbp]
\centering
\begin{tabular}{cc}
\includegraphics[height=6cm]{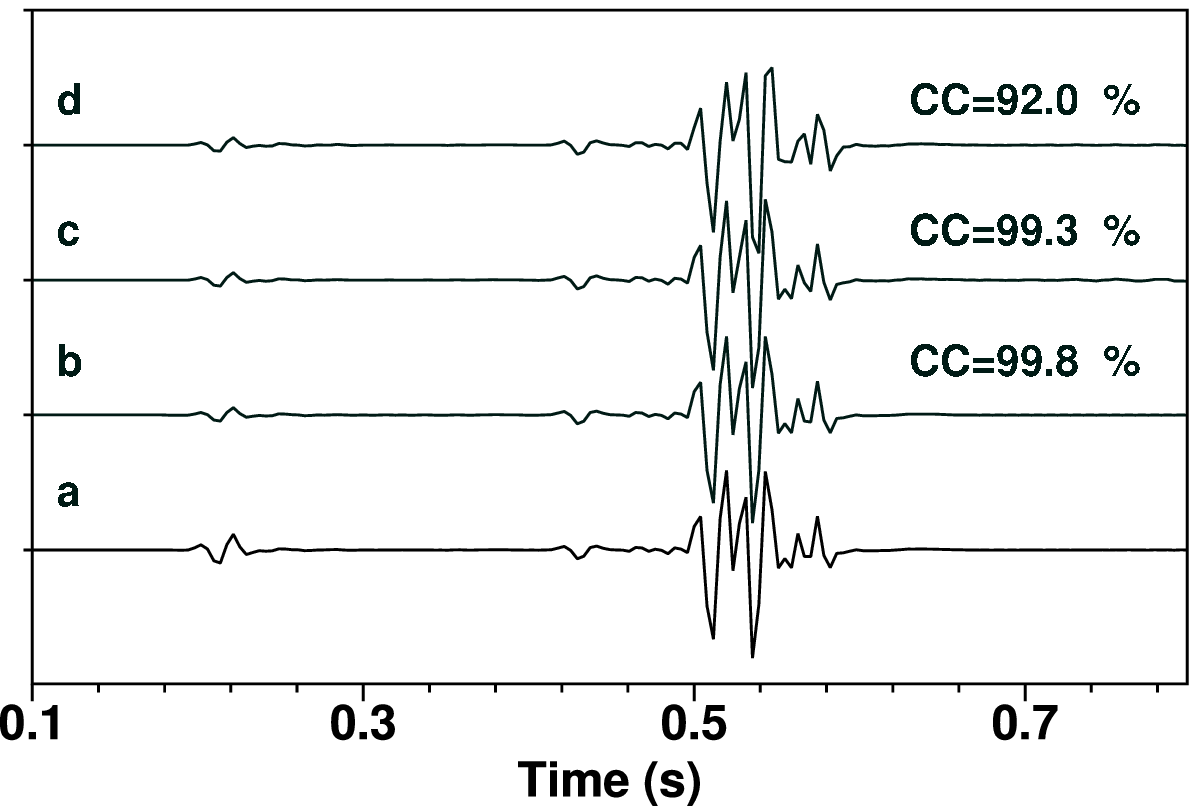} & 
\end{tabular}
\caption{Detailed comparison of individual seismic traces obtained with the Fr\'echet derivative and with the discrete perturbation methods for various perturbation amplitudes. The model used for the computations is the 16-layer model presented in figure \ref{fig2}. The seismic source is a vertical point force. The figure shows the perturbation of the vertical displacement with respect to the consolidation parameter $cs$ for a source-receiver offset of 500 m. The seismogram labeled {\em a} corresponds to the Fr\'echet derivative formulation. The traces labeled {\em b}, {\em c} and {\em d} are respectively associated with discrete perturbations of the model parameter $cs$ with relative amplitude contrasts of 1 \%, 10 \% and 50 \%. The similarity between the Fr\'echet derivative and discrete perturbation seismograms is quantified by their correlation coefficient which is indicated above the traces.}
\label{fig4}
\end{figure}

Figure \ref{fig4} shows that the Fr\'echet derivative seismograms are remarkably accurate as long as the perturbation amplitude remains small. 
In this case, the maximum acceptable perturbation amplitude is approximatively 10 \% of the model parameter value. 
When the perturbation amplitude is increased beyond this limit, the $P-$waveforms remain practically unchanged whereas the $S-$waveforms are distorted. 
In all cases, no variations in travel times are observed.

We also checked the behavior of the Fr\'echet derivative operators relative to the thickness of the perturbation layer, or equivalently, relative to the central frequency of the wavelet used in the computations. 
As for the uniform medium investigated before, we note that the first-order approximations remain very accurate as long as the layer thickness layer is lower than $\lambda_P/5$ or $\lambda_S/5$ depending on the model parameter considered. 
   
\section{Sensitivity study}

\subsection{Relative influence of the model parameters}

\begin{figure}[!hbp]  
\centering
\includegraphics[height=10cm]{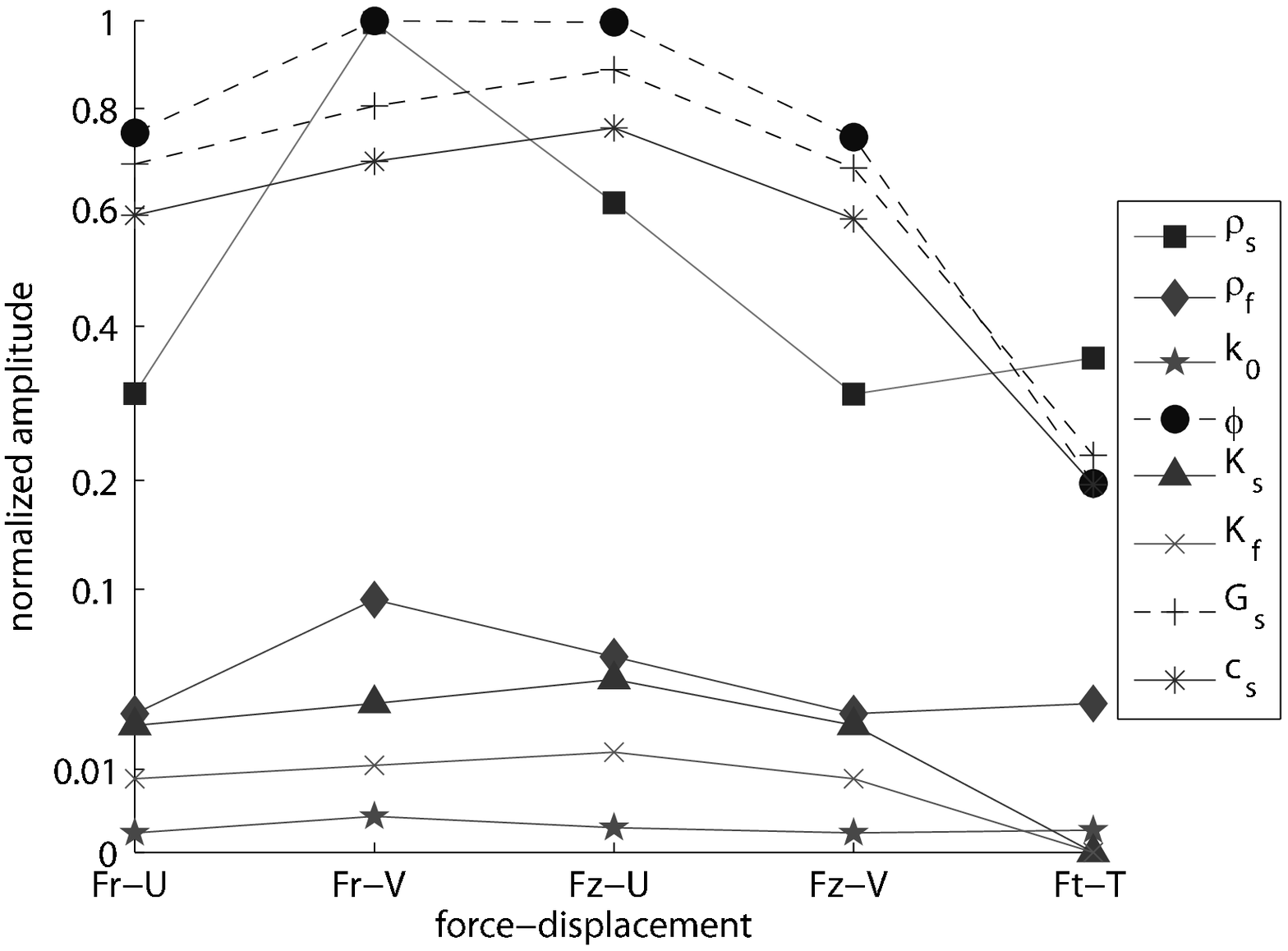}  
\caption{Normalized maximum amplitude of the Fr\'echet derivative seismograms for all solid force and displacement pairs. $F_z$, $F_r$ and $F_t$ respectively denote the vertical, horizontal radial and horizontal tangential components of the forces acting on an average volume of porous medium. $U$, $V$ and $T$ stand for the vertical, horizontal radial and horizontal tangential components of the solid displacement. The model parameters are those given in table 1 except for the consolidation parameter $c_s$ which is equal to 20. The amplitude scale is non linear for sake of readability.} 
\label{fig5}
\end{figure}
 
\begin{figure}[!hbp]  
\centering
\includegraphics[height=10cm]{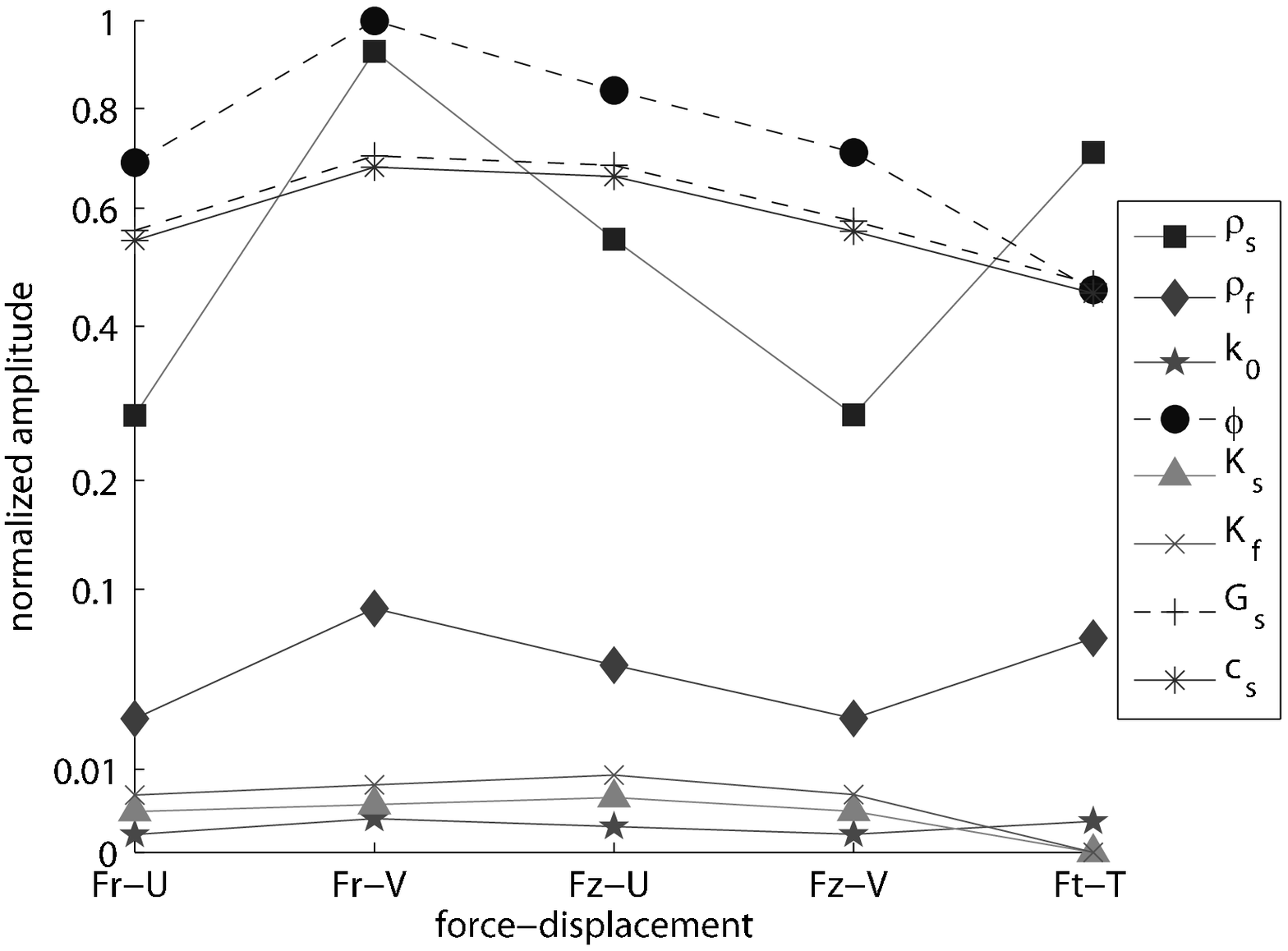}
\caption{Same as figure \ref{fig5} with consolidation parameter $c_s$ equal to 100.} 
\label{fig6}
\end{figure} 

In this section, we assess the relative influence of small modifications of the model parameters on the different components of the seismic wave field.
For this, we determine the maximum amplitude of the seismograms computed with the discrete perturbation method for all pairs of vertical and horizontal forces and displacements.
These computations are done with the model parameters of table 1, aside from the
consolidation parameter $c_s$ which is given a value of 20 in figure \ref{fig5} and 100 in figure \ref{fig6}. 
Perturbation depth, source and receiver locations and maximum offset are identical to those used in section IV.B.
The amplitudes thus obtained are multiplied by the parameter variation ($\Delta p$) to obtain  the displacement change $\Delta U$ of equation (\ref{eq401}), and are normalized with respect to the maximum value found. 
The same quantities were computed with the Fr\'echet derivative approach to check the agreement between the two computation techniques. 
Figures \ref{fig5} and \ref{fig6} show that the seismograms are essentially sensitive to porosity $\phi$ in the uniform medium considered. 
The seismograms are also strongly influenced by perturbations of the consolidation parameter $c_s$, mineral density $\rho_s$ and shear modulus $G_s$. 
On the other hand, changes in fluid density $\rho_f$, mineral modulus $K_s$, fluid modulus $K_f$ and permeability $k_0$ have only a weak influence on the wave amplitudes. 
We also note that the influence of parameters $c_s$ and $G_s$ on the one hand, and $\rho_s$ and $\rho_f$ on the other hand, are very similar for the various force-displacement pairs.

For the model with the lowest value of the consolidation parameter (corresponding to the most consolidated material, figure \ref{fig5}), the solid modulus $K_s$ has a stronger influence on the seismograms than the fluid modulus $K_f$. 
On the contrary, for an unconsolidated medium (figure \ref{fig6}), the seismograms are mainly influenced by the fluid properties. 
With the model parameters used in this study, we find that the transition between these two regimes occurs for a consolidation parameter of 35. 
We also verified, as suggested by our observations in section IV.B, that the porosity parameter shows the same behavior: the seismogram amplitudes in a high porosity medium depend more strongly on the fluid modulus than on the solid modulus, and vice-versa for a medium with low porosity. 
As a consequence (and confirmation of field observations), the $P$-waves are strongly influenced by the fluid properties when they propagate in a fluid-saturated and poorly consolidated medium with high porosity.
This makes it possible to determine the fluid characteristics from the seismic waveforms if these favorable conditions are met. 
Conversely, the estimation of the fluid characteristics will be more difficult in consolidated or unsaturated or low porosity media. 
Another consequence of the results shown in figures \ref{fig5} and \ref{fig6} is that $\phi$ and $c_s$ are the most attractive parameters to invert for in an inversion procedure, assuming that $\rho_s$ and $G_s$ can be estimated independently. 

\subsection{Amplitude of the perturbation seismograms versus angle of incidence}

\begin{figure}[!hbp]  
\begin{center}
\includegraphics[height=18cm]{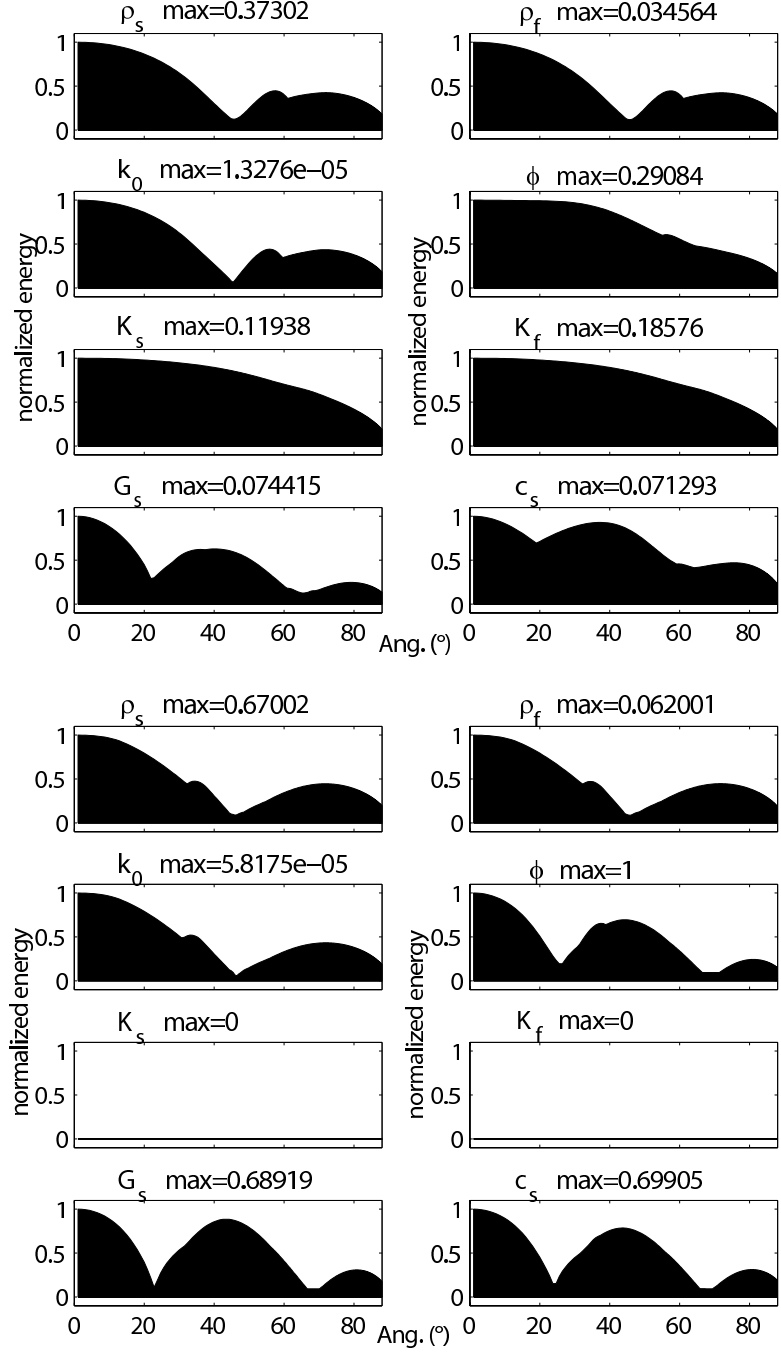}
  % pap.s2_Vp.eps: 1048592x1048592 pixel, 0dpi, infxinf cm, bb=
\end{center}
% \begin{center}
% \includegraphics[height=9cm]{figure/fig7b.eps}
% \end{center} 
\caption{Energy of plane waves reflected from perturbations in $\rho_s$, $\rho_f$, $k_0$, $\phi$, $K_s$, $K_f$, $G_s$ and $cs$, as a function of incidence angle. The eight upper panels and eight lower panels respectively correspond to $PP$ and $SS$ reflections. The curves are normalized with respect to the maximum value indicated above each panel.}
\label{fig7}
\end{figure} 

The previous figures already stressed the interdependence (or coupling) of some model parameters.
Parameter coupling means that small perturbations of two or more parameters result in similar modifications of the seismic response.
An obvious consequence of parameter coupling is that it becomes difficult or even impossible to reliably estimate the model parameters in an inversion procedure \citep{tarantola86}. 

To look into this problem, we computed the plane wave responses corresponding to 10\% perturbations of the model properties in the infinite medium described in table 1.
Figure \ref{fig7} shows the reflected energy for the eight model parameters as a function of angle of incidence at the perturbed layer, both for the $PP$ and $SS$ reflections.
The smooth aspect of the curves is due to the intrinsic attenuation of the seismic waves propagating in the porous medium.
The peaks and troughs of the curves are explained by the strong variations of the reflection and transmission coefficients, as shown for instance by \citet{delacruz92}. 
Some of these rapid variations are seen on the seismograms of figure \ref{fig1} which were computed with the same model and source-receiver configuration. 

The magnitudes of the seismic responses shown in figure \ref{fig7} are consistent with the study presented in section V.A. 
$SS$ reflections are about twice as large as $PP$ reflections for perturbations of the solid and fluid densities. 
For $SS$ reflections, the maximum value is reached for a perturbation in porosity $\phi$, whereas $G_s$, $c_s$ and $\rho_s$ produce perturbations of the same magnitude. 
For $PP$ reflections, the most influential parameters are the solid density $\rho_s$ and porosity $\phi$.

We also see in figure \ref{fig7} that the radiation patterns associated with perturbations in $\rho_s$, $\rho_f$ and $k_0$ on the one hand, and $K_s$ and $K_f$ on the other hand are exactly the same for the $PP$ reflections. 
The same resemblance is observed for the $SS$ reflections for the $\rho_s$, $\rho_f$ and $k_0$  group of parameters, and for the $G_s$, $\phi$ and $c_s$ group of parameters. 
We note in particular that permeability $k_0$ and densities $\rho_s$ and $\rho_f$ behave similarly despite their different roles in the constitutive equations. 
In all cases, the backscattered energy is maximum at normal incidence because of the shorter wave path and corresponding minimal attenuation. 

\section{Conclusions}

We derived the Fr\'echet derivatives of the seismic response of a depth-dependent porous medium. 
The Fr\'echet derivatives are analytically expressed in terms of the Green's functions of the propagation medium through a perturbation analysis of the poro-elastic wave equations expressed in the plane wave domain. 
Started with a primary set of seven model parameters chosen because of their linear relationship with the wave equations, the derivation was carried on with a secondary set of eight model parameters more convenient to use as physical parameters of the problem.
The eight model parameters considered in our analysis are related to the fluid properties (density, bulk modulus), to the mineral properties (density, bulk modulus, shear modulus) and to the arrangement of the porous material (porosity, permeability and consolidation parameter). In the $P-SV$ case, we derived Fr\'echet derivatives for 3 different sources (horizontal and vertical point forces and explosive point source), 4 displacement components and 8 model parameters. 
In the $SH$ case, we obtained 12 expressions for 1 horizontal point force, 2 displacement components and 6 model parameters leading to non-zero Fr\'echet derivatives. \\
We checked the accuracy of these sensitivity operators in the time-distance domain by comparing the waveforms computed from the first-order expressions with seismograms obtained by introducing discrete perturbations in the medium properties.
The numerical tests were carried out both in a homogeneous and in a more complex earth model excited by oriented point forces.
By and large, we found that our analytical expressions of the Fr\'echet derivatives are remarkably accurate as long as the Born approximation assumptions are satisfied, that is, as long as the perturbations of the model parameters are weak and localized. 
However, as in other studies relying on the Born approximation, we showed that the first-order operators are robust enough to model parameter perturbations up to 20 \% and layer thicknesses up to one fifth of the dominant wavelength.  Furthermore, our formulation appears to be stable at all source-receiver offsets.\\
Due to their analytical formulation, the sensitivity operators derived in this paper will be especially useful in full waveform inversion algorithms implemented with gradients techniques. As a first step toward such an application, we evaluated the sensitivity of the seismic response of a poro-elastic medium with respect to each model parameter. 
We showed that the porosity and consolidation parameter are the most attractive parameters to invert for, whereas the permeability appears to be the most difficult parameter to determine. The wave fields are more sensitive to the fluid bulk modulus than to the mineral bulk modulus, or inversely, according to porosity and consolidation parameter values. A multi-parameter inversion of backscattered energy looks challenging because of the strong coupling of several model parameters in a wide range of angles of incidence.
Finally, this sensitivity study should prove useful for the interpretation of time-lapse monitoring surveys and for checking solutions (yet to come) accounting for 3D heterogeneities in the propagation medium.

\begin{acknowledgments}

We are grateful to St\'ephane Garambois, Michel Bouchon, Jean-Louis Auriault, Patrick Rasolofosaon and Helle Pedersen for many helpful discussions in the course of this work. 
The numerical applications were performed by using the computer facilities of the Grenoble Observatory. 
We thank the Associate Editor and two anonymous reviewers for their useful comments.
\end{acknowledgments}
%\newpage
%--------------------------------------------- bibliographie ----------------------------------------------- 

\newpage
%\listoffigures

\end{document}